# A Bidirectional Reflectance Distribution Function for VisorSat Calibrated with 10,628 Magnitudes from the MMT-9 Database

Anthony Mallama

anthony.mallama@gmail.com

2021 September 15

**Abstract**

A BRDF for the VisorSat model of Starlink satellites is described. The parameter coefficients were determined by least squares fitting to more than 10,000 magnitudes recorded by the MMT-9 robotic observatory. The BRDF is defined in a satellite-centered coordinate system (SCCS) which corresponds to the physical shape of the spacecraft and to the direction of the Sun. The three parameters of the model in the SCCS are the elevations of the Sun and of MMT-9 along with the azimuth of MMT-9 relative to that of the Sun. The mean VisorSat magnitude at a standardized distance of 1,000 km is 6.84 and the RMS of the distribution around that mean is 1.05. After the magnitudes are adjusted with the BRDF, the RMS reduces to 0.51. The set of MMT-9 observations transformed to the SCCS is available from the author.

## 1. Introduction

The population of bright satellites in low-earth-orbit is growing rapidly. The number is already so large that the spacecrafts are interfering with ground-based astronomy (Walker et al., 2020a, Walker et al., 2020b, Tyson et al., 2020, Otarola et al., 2020, Gallozzi et al., 2020, Hainaut and Williams, 2020, McDowell, 2020, Williams et al. 2021a, Williams et al. 2021b, Lawler et al., 2021, Mallama and Young, 2021). More than a thousand Starlink communication satellites have been launched by the SpaceX company and many more are planned. OneWeb is actively pursuing a similar project and other companies and countries may follow.

SpaceX is cooperating with the astronomical community and has developed a VisorSat model of Starlink that includes a Sun shade to make the spacecraft appear fainter. Mallama (2021) determined the mean brightness of these new spacecraft in visible light based on 720 observed magnitudes. The study reported here characterizes VisorSat brightness in more detail using a bidirectional reflectance distribution function (BRDF) which is based on 10,628 magnitudes.

Section 2 describes the MMT-9 automated observatory which is the source of the observed magnitudes used to calibrate the BRDF. Section 3 defines a satellite-centered coordinate system (SCCS) that corresponds to the shape of a Starlink spacecraft. The SCCS is the frame of reference for the BRDF. Section 4 describes the processing of data retrieved from the MMT-9 database, including determination of the positions of the Sun and the MMT-9 in the SCCS system. Section 5 explains how the parameter coefficients of the BRDF were evaluated and illustrates how the model reduces magnitude uncertainties. Section 6 discusses this research in the context of other studies of Starlink satellite brightness. Section 7 cites the limitations of this study. The conclusions are given in Section 8.

## 2. MMT-9 system, magnitudes and database

Mini-MegaTORTORA (MMT-9) is a robotic observatory (Karpov et al. 2015) in Russia. The MMT-9 hardware includes nine 71 mm diameter f/1.2 lenses and 2160 x 2560 sCMOS sensors. The detectors are sensitive to the visible spectrum from red through blue. A color transformation formula kindly provided by S. Karpov (private communication) indicates that MMT-9 magnitudes taken through the clear filter are within 0.1 magnitude of the Johnson-Cousins V band-pass for objects with small B-V color indices.

The [MMT-9 database](#) contains observations of Starlink satellite passes that are collected into *track* files. The author downloaded 26 VisorSat track files which contain more than 10,000 records. The satellite magnitudes were recorded between January 3 and May 8 of 2021. All spacecraft were at the operational altitude (approximately 550 km) according to the [plots](#) created by J. McDowell.

The program [JSatTrak](#) was used to generate ground-based azimuth and elevation values for the VisorSat at the moment when each MMT-9 observation occurred. The sky coverage is illustrated in Figure 1 where azimuth is that of the satellite relative to that of the Sun. The distribution of VisorSat positions above elevation $30^o$ is reasonably uniform except in the direction of the Sun where there are no observations below $45^o$.

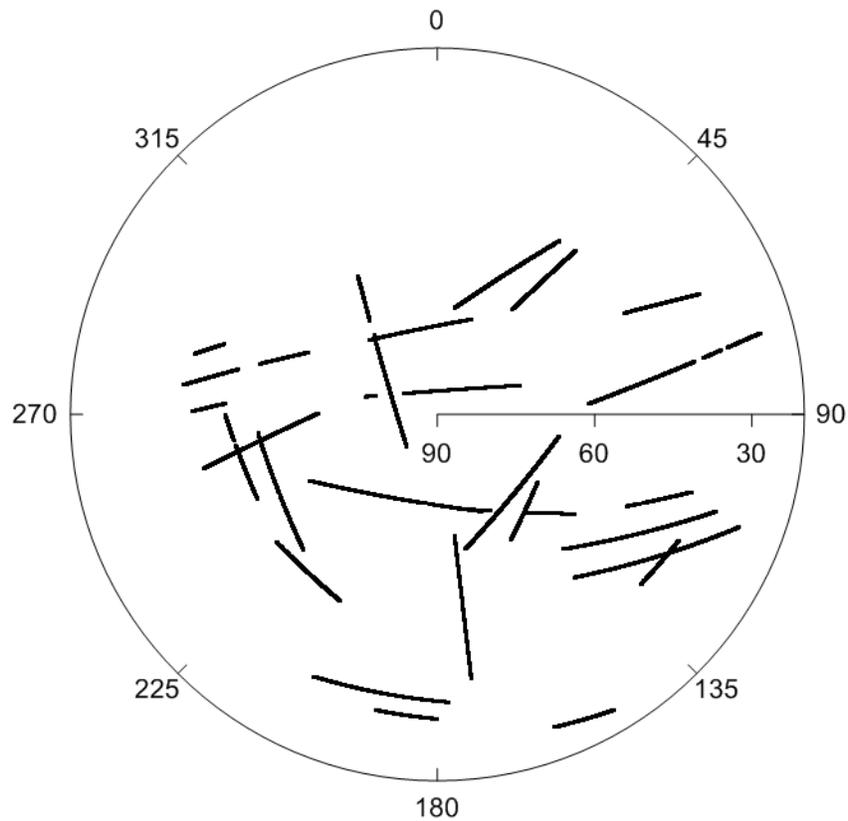

Figure 1. The VisorSats tracks corresponding to the data used in this study are plotted in the sky above the MMT-9 site as azimuth relative to the Sun and as elevation above the horizon.

## 3. Satellite-centered coordinate system

The largest physical components of Starlink spacecrafts are a flat-panel shaped body and a solar array. When the satellites are at their operational altitude the body panel is nominally perpendicular to the nadir direction and the solar array is nominally perpendicular to the solar direction.

This geometry suggests a satellite-centered coordinate system (SCCS) for BRDF modeling. The north pole of the SCCS is toward the satellite zenith. The important directions for any BRDF are those of the lighting source and of the sensor which, in this case, correspond to the Sun and the MMT-9 observatory. The components of the direction vector are azimuth and elevation. The azimuth of the SCCS is defined here to be zero in the direction of the Sun. So, the azimuth of MMT-9 is its offset from that of the solar direction. The elevations of the Sun and of MMT-9 are measured relative to the north pole of the SCCS. Azimuths and elevations in the SCCS are distinct from those in the ground-based coordinate system. The SCCS geometry is illustrated in Figure 2.

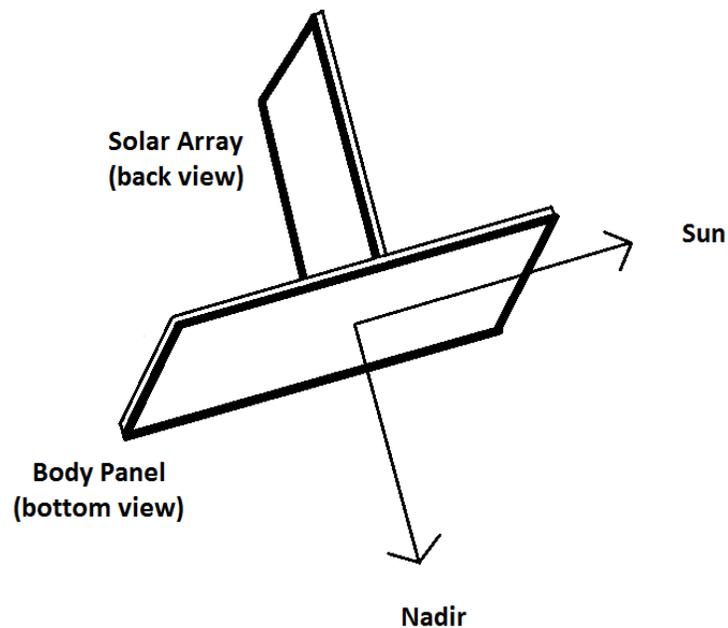

Figure 2. This schematic diagram of a Starlink spacecraft illustrates the SCCS frame of reference. The north pole of the SCCS is in the direction opposite from nadir. The body panel is nominally in the plane of the SCCS equator. Most solar and MMT-9 elevations are negative when the satellites are visible from the ground. The zero point of azimuth is in the solar direction.

## 4. Data processing

The coordinates of the MMT-9 site and of the Sun in the SCCS were computed as follows. Two line element sets (TLEs) corresponding to the spacecraft orbit at the time of observation were obtained from Space-Track. The spacecraft XYZ coordinates in the Earth-Centered Earth-Fixed reference frame were then generated by the JSatTrak program. The corresponding right ascension (RA) and declination (Dec) values for the north pole of the SCCS were adjusted for precession to the time of the observation. The XYZ positions of the MMT-9 site were calculated from its latitude, longitude and height, along with the sidereal time. The precession-adjusted XYZ coordinates of the satellite were subtracted from those of MMT-9 to determine the RA and Dec of the site which, in turn, were transformed to its azimuth and elevation in the SCCS. The solar RA and Dec and the sidereal time were obtained from the JPL Horizons on-line ephemeris. The solar coordinates were transformed to azimuth and elevation in the SCCS. Finally, the MMT-9 azimuth offset was calculated by subtracting the solar azimuth.

The MMT-9 track files report apparent magnitudes. Those were standardized by adjusting them to a range of 1000 km.

## 5. BRDF calibration

Coefficients of the three BRDF parameters (solar and MMT-9 elevation and MMT-9 azimuth) were determined by fitting them to the 1,000-km magnitudes. Brightness was found to be highly sensitive to the MMT-9 azimuth as shown by the quadratic fit in Figure 3.

The next most sensitive parameter, after azimuth, is solar elevation which is represented by a linear fitting to the residuals from the azimuth fit. Finally, MMT-9 elevation was fit linearly to the magnitude residuals from the combined fit of azimuth and solar elevation. The coefficients from this fitting process are listed in Table 1.

Figure 4 illustrates the application of the BRDF to the 1000-km magnitudes. The mean magnitude at that standard distance is 6.84 and the root mean square of the differences of individual observations from that mean is 1.05. When those magnitudes are adjusted with the BRDF fitting, the RMS residual reduces to 0.53. After one iteration of the fitting, the RMS is further reduced to 0.51.

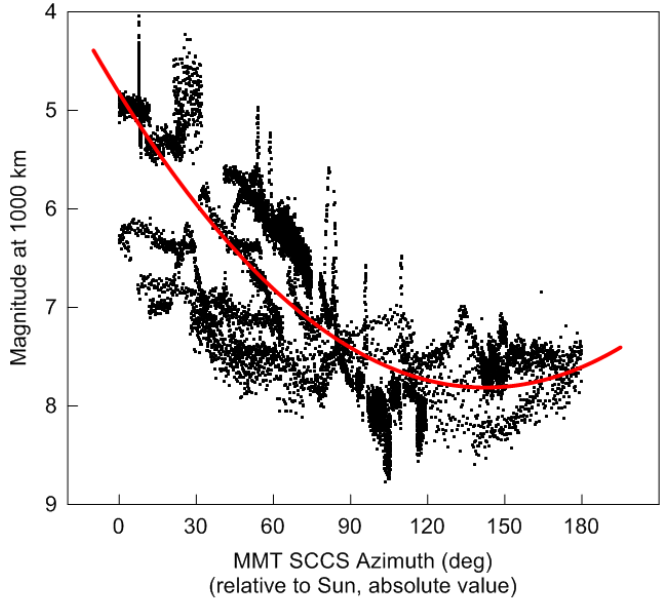

Figure 3. VisorSat is brightest when the SCCS azimuths of the Sun and MMT-9 are aligned. The dimming trend levels off beyond 90°.

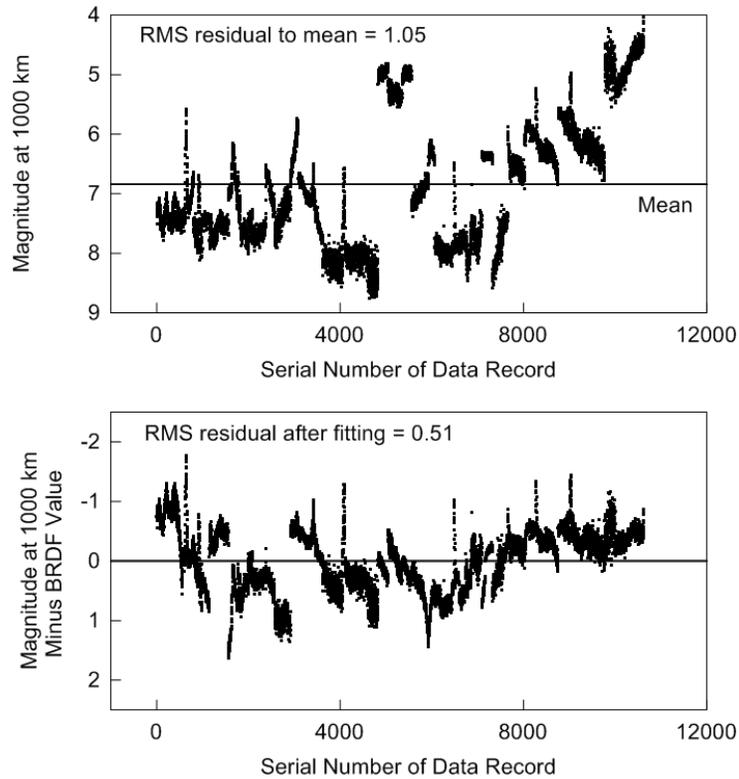

Figure 4. The scatter of magnitudes is reduced by more than half when the BRDF is applied.

Table 1. Polynomial coefficients of the BRDF parameters after 1 iteration

```
                   -------- Coefficient -------
Parameter             0         1         2
---------          --------  --------  --------
Azimuth            4.935479  0.042445 -0.000167
Solar elevation    0.986733  0.059549
MMT elevation     -0.940909 -0.016837

Coefficients apply to angles in degrees
```

## 6. Discussion

The brightness of Starlink satellites has been addressed in several previous studies. Otarola et al. (2020) and Tregloan-Reed et al. (2020) have reported the results of filtered photometry. Lawler et al. (2021) presented g'-band results. Mallama (2020b and 2021) assessed the visual brightness of the original design of Starlink spacecraft and then compared that to the reduced brightness of VisorSat.

Cole (2020 and 2021) has developed a BRDF model for VisorSat which represents the shape and the reflecting properties of the spacecraft body and of the solar array in detail. Mallama (2020a) reported on a BRDF model for the Starlink spacecraft body but it did not solve for the azimuth parameter.

While few magnitude observations of Starlink satellites have been published, observing programs are underway. Likewise, new BRDF studies are being pursued. The astronomical community is also planning for a central repository of satellite observations and models.

## 7. Limitations of this study

Several factors place limits on the general applicability and the accuracy of the results from this study. First is that the satellite body is only *nominally* aligned with the SCCS, while its *actual* attitude (yaw, pitch and roll) at the time of each observation is unknown. Thus, the presumed alignment between angles in the SCCS and the BRDF parameters corresponding to the physical spacecraft is only approximate.

The data are all from one location. If spacecraft attitude is being adjusted geographically for operational reasons then the magnitudes acquired by MMT-9 in Russia may not be representative of other regions.

The observations only sample a 5 month time span. Cole (2021) found that VisorSats viewed in the anti-Sun direction had become brighter in 2021. This change may be due to an adjustment of the satellite attitude or may be caused by a change in the angle of the solar array to the spacecraft body. In any case, the results presented here may not represent other time periods very well.

There is limited physical modeling of the spacecraft. The body panel is taken to be in the plane of the SCCS, which motivates fitting the elevations of the Sun and the MMT-9 site. Likewise, the solar array is taken to be in the direction of the Sun which motivates fitting the azimuth difference between the Sun and the MMT-9 site. However, there is no modeling for the Sun shade or any of the other physical spacecraft attributes. Likewise, no information about the light scattering properties of the satellites surfaces is included.

There is moderate scatter in the MMT-9 magnitudes. There are 2,754 instances of time-adjacent observations in the MMT-9 track files where the same satellite was recorded by different instrument channels within one second of time. The RMS difference of those observations is 0.19 magnitude.

The data set analyzed here is not entirely uniform in sky coverage. In particular, there is no data for satellites that were located toward the solar azimuth and below 45 degrees elevation in the sky.

## 8. Conclusions

A set of 10,628 magnitudes of VisorSat satellites recorded by the MMT-9 robotic observatory has been supplemented with information needed for BRDF modeling. This additional information includes elevation and azimuth angles in a satellite-centered coordinate system, as well as magnitudes adjusted to a standard distance of 1,000 km. The BRDF developed in this study reduces the RMS scatter of VisorSat 1000-km magnitudes by more than half.

**Acknowledgements**

## References

Cole, R.E. 2020. A sky brightness model for the Starlink "Visorsat" spacecraft. Research Note of the AAS. https://iopscience.iop.org/article/10.3847/2515-5172/abc0e9.

Cole, R.E. 2021. A sky brightness model for the Starlink "Visorsat" spacecraft. https://arxiv.org/abs/2107.06026.

Gallozzi, S., Scardia, M., and Maris, M. 2020. Concerns about ground based astronomical observations: a step to safeguard the astronomical sky. https://arxiv.org/pdf/2001.10952.pdf.

Hainaut, O.R., and Williams, A.P. 2020. Impact of satellite constellations on astronomical observations with ESO telescopes in the visible and infrared domains. *Astron. Astrophys.* manuscript no. SatConst. https://arxiv.org/abs/2003.019pdf.

Karpov, S., Katkova, E., Beskin, G., Biryukov, A., Bondar, S., Davydov, E., Perkov, A. and Sasyuk, V. 2015. Massive photometry of low-altitude artificial satellites on minimegaTORTORA. Fourth Workshop on Robotic Autonomous Observatories. RevMexAA.

Lawler, S.M., Boley, A.C. and Rein, H. 2021. Visibility predictions for near-future satellite megaconstellations: Latitudes near $50^o$ will experience the worst light pollution. https://arxiv.org/abs/2109.04328.

Mallama, A. 2020a. A flat-panel brightness model for the Starlink satellites and measurement of their absolute visual magnitude. https://arxiv.org/abs/2003.07805.

Mallama, A. 2020b. Starlink satellite brightness before VisorSat. https://arxiv.org/abs/2006.08422.

Mallama, A. 2021. The Brightness of VisorSat-Design Starlink Satellites. https://arxiv.org/abs/2101.00374.

Mallama, A. and Young, M. 2021. Beyond Starlink: the satellite saga continues. Sky and Telescope, Vol. 141, June 2021, pp. 16-19.

McDowell, J. 2020. The low Earth orbit satellite population and impacts of the SpaceX Starlink constellation. *ApJ Let*, 892, L36 and https://arxiv.org/abs/2003.07446.

Otarola, A. (chairman) and Allen, L., Pearce, E., Krantz, H.R., Storrie-Lombardi, L., Tregloan-Reed, J, Unda-Sanzana, E., Walker, C. and Zamora, O. 2020. Draft Report of the Satellite Observations Working Group commissioned by the United Nations, Spain and the International Astronomical Union as the Workshop on Dark and Quiet Skies for Science and Society. https://owncloud.iac.es/index.php/s/WcdR7Z8GeqfRWxG#pdfviewer.

Tregloan-Reed, J, Otarola, A., Ortiz, E., Molina, V., Anais, J., Gonzalez, R., Colque, J.P. and Unda-Sanza, E. 2020. First observations and magnitude measurement of SpaceX's Darksat. *Astron. Astrophys.*, manuscript no. Darksat_Letter_arXiv_submission_V2. https://arxiv.org/pdf/2003.07251.pdf.

Tyson, J.A., Ivezić, Ž., Bradshaw, A., Rawls, M.L., Xin, B., Yoachim, P., Parejko, J., Greene, J., Sholl, M., Abbott, T.M.C., and Polin, D. (2020). Mitigation of LEO satellite brightness and trail effects on the Rubin Observatory LSST. arXiv e-prints, arXiv:2006.12417.

Walker, C., Hall, J., Allen, L., Green, R., Seitzer, P., Tyson, T., Bauer, A., Krafton, K., Lowenthal, J., Parriott, J., Puxley, P., Abbott, T., Bakos, G., Barentine, J., Bassa, C., Blakeslee, J., Bradshaw, A., Cooke, J., Devost, D., Galadí-Enríquez, D., Haase, F., Hainaut, O., Heathcote, S., Jah, M., Krantz, H., Kucharski, D., McDowell, J.loan-Reed, J., Wainscoat, R., Williams, A., and Yoachim, P. 2020a. Impact of satellite
9